\documentclass[journal,12pt,onecolumn,draftclsnofoot,]{IEEEtran}
\usepackage{graphicx}
\usepackage{float}
\usepackage{amsmath}
\usepackage{cases}
\usepackage{amsfonts}
\usepackage{algorithm}
\usepackage{color}
\usepackage{algorithmic}
\usepackage{cite}
\begin{document}

\title{Characterizing Energy Efficiency of Wireless Transmission for Green Internet of Things: A Data-Oriented Approach}

\author{
Hong-Chuan Yang, \textit{Senior Member, IEEE} and Mohamed-Slim Alouini, \textit{Fellow, IEEE}
\thanks{This work was supported in part by an NSERC Discovery Grant.}
\thanks{H.-C. Yang is with the Department of Electrical and Computer Engineering, University of Victoria, Victoria, BC V8W 2Y2, Canada (e-mail: hy@uvic,ca).}
\thanks{M.-S. Alouini is with the
Computer, Electrical, and Mathematical Sciences and Engineering (CEMSE) Division, King Abdullah University of Science
and Technology (KAUST), Thuwal 23955, Saudi Arabia (e-mail: slim.alouini@kaust.edu.sa).}
}


\IEEEtitleabstractindextext{
\begin{abstract}
The growing popularity of Internet of Things (IoT) applications brings new challenges to the wireless communication community. Numerous smart devices and sensors within IoT will generate a massive amount of short data packets. Future wireless transmission systems need to support the reliable transmission of such small data with extremely high energy efficiency. In this article, we introduce a novel data-oriented approach for characterizing the energy efficiency of wireless transmission strategies for IoT applications. Specifically, we present new energy efficiency performance limits targeting at individual data transmission sessions. Through preliminary analysis on two channel-adaptive transmission strategies, we develop several important design guidelines on green transmission of small data. We also present several promising future applications of the proposed data-oriented energy efficiency characterization. 
\end{abstract}

}

\maketitle

\IEEEdisplaynontitleabstractindextext
\IEEEpeerreviewmaketitle

\section{Introduction}

\IEEEPARstart{T}{he} Internet of Things (IoT) will dramatically change the way we interact with the world. IoT extends the Internet to our daily objects, such as appliances, cameras, lights, displays, and vehicles, by equipping them with micro-controllers, communication, and networking capability. Such extension will transform our daily life and enable many new applications, ranging from home automation and traffic management, to smart grids and mobile health-care \cite{7809016}. Furthermore, these connected devices will generate a large amount of data, the timely processing of which will bring huge social and economical benefits. Meanwhile, several technical challenges need to be addressed to realize the full potential of IoT. For example, the IoT needs to support diverse application scenarios, which typically have diverse service requirements \cite{6740844}. The provision of IoT functionality will definitely increase the overall system cost, the justification of which requires suitable business model. Furthermore, the communication and networking functions of IoT devices will necessarily consume extra energy. As the number of connected devices and sensors  within the IoT will be enormous, the overall energy consumption of future IoT could be prohibitive with conventional transmission strategies. As such, there is a pressing need for developing green IoT technologies. 

Wireless transmission is the idea choice for connecting IoT devices. Therefore, designing highly energy efficient wireless transmission strategies will be essential to the realization of green IoT. Energy efficiency has always been a serious concern for wireless systems since wireless devices typically have limited energy supply. Various advanced transmission technologies, including 
channel adaptive transmission \cite{chua98} and cooperative relay transmission\cite{sendonaris1, laneman2}, are developed and deployed to support high data rate wireless services with low energy cost. These transmission technologies were typically designed with the goal of enhancing or approaching the capacity limits of wireless channels for a given transmission power, as the energy efficiency is usually quantified as the ratio of channel capacity over transmission power \cite{5783982, alsharif_JCNC13}. On the other hand, most existing metrics characterize energy efficiency in an average sense. Such characterization can not provide useful guidelines to the energy efficiency improvement for individual transmission sessions over IoT, which usually occur in a sporadic fashion.

The IoT introduces a paradigm shift to wireless communications. Most IoT applications entails quick information exchanges from smart devices/sensors.  These machine-type terminals will sporadically access the networks for the transmission of short packets that contains metering data, status information, and remote commands. These transmission sessions will have much shorter duration than conventional traffics,  such as phone calls and video streaming. Conventional transmission system design typically adopts a {\emph channel-oriented} approach assuming a consistent and continuous data traffic and improves the average channel quality with advanced transmission technologies. Meanwhile, such approach ignores the specifics of individual transmission sessions, such as the traffic characteristics and the prevailing channel/network condition. When the transmission sessions are short, the energy efficiency achieved by individual sessions will vary dramatically as the result of the channel variation. To further improve the energy efficiency of wireless transmission systems, especially for IoT applications, we need to optimally design the  transmission strategies from the perspective of individual transmission sessions. 

Motivated by this intuition, we propose a novel \emph{data-oriented} approach for the energy efficiency optimization of wireless transmission strategies for IoT applications. Specifically, when a certain amount of data is available for transmission, we optimally decide the transmission strategy for the  highest possible energy efficiency. The transmission strategy will be adjusted for each data transmission session according to the traffic characteristics and the channel/network conditions. Intuitively, we expect that the average energy efficiency of wireless transmission will be further enhanced if the transmission strategy is optimized for each transmission session. In this article, we present an initial investigation on this data-oriented approach for developing energy efficient transmission strategies. In particular, we introduce two new data-oriented energy efficiency performance metrics targeting at individual data transmission sessions. We illustrate their analysis on two popular channel adaptive transmission strategies over fading channels. Finally, we discuss several promising future research directions with the data-oriented approach for green wireless transmission system design and analysis. 




\section{Conventional energy efficiency metrics}

Energy efficiency metrics are essential to the analysis and design of green communication systems.  They help assess and compare the energy consumption of different designs and provide long-term research goals. The energy efficiency metrics for wireless communication systems can be generally classified into two categories: i) network-level metrics and ii) link-level metrics. Network-level metrics characterize the energy efficiency of the whole system with the consideration of service coverage. Examples include the ratio of coverage area to site power consumption with the unit of km$^2$/Watt \cite{etsi09}  and the average power usage per service data rate per coverage area in Watt/bps/m$^2$ \cite{5379031}. As many factors, including equipment choice, network structure, and facility arrangement, affect the energy consumption of a wireless network, these network-level metrics can not provide direct guidelines to green design of wireless transmission system.

Link-level metrics typically focus on the energy efficiency of a particular transmission link and quantify the efficiency of the transmission system in achieving a certain transmission rate with respect to resource utilization. For example,  the achieved data rate per unit power consumption, with unit of bits/s/Watt or equivalently bits/Joule, is a widely used energy efficiency metric\cite{alsharif_JCNC13}. This metric was applied to the tradeoff analysis among different system design parameters \cite{5783982}. The radio efficiency metric in m$\cdot$bit/s/Watt \cite{5956139} considers both transmission rate and transmission distance. With the application of Shannon capacity formula, the upper bounds of these energy efficiency metrics can be evaluated. On the other hand, these metrics are typically defined for constant channel realization with fixed transmission power and as such can not directly apply to fading wireless channels with time-varying channel gains.

We can generalize most link-level metrics to fading wireless channels by applying the ergodic capacity concept. Ergodic capacity characterizes the largest possible average transmission rate that a wireless channel can support. Using ergodic capacity, we can evaluate the average energy efficiency of wireless transmission over fading channels. In particular, the ergodic capacity was utilized to evaluate the area spectral efficiency of cellular systems\cite{600409}. The metric was later generalized to quantify the energy efficiency of point-to-point transmission with the consideration of affected area \cite{6712192}. Meanwhile, these ergodic capacity based metrics on can only characterize the energy efficiency of wireless transmission in an average sense. The resulting analysis is generally applicable to conventional continuous data traffics. The IoT involves numerous machine-type terminals that generates sporadic small data packets. The  energy consumption of individual data transmission session for these small data varies dramatically with the prevailing channel realization. The realization of green IoT relies heavily on the energy efficiency improvement for short transmission sessions. 

To further enhance the energy efficiency of wireless system for `small data' transmission, we need to study wireless transmission technologies from a new perspective. In this article, we follow a data-oriented approach and propose to characterize the energy efficiency of wireless transmission for the perspective of individual data transmission sessions. More specifically, we raise the following fundamental questions: Given a certain amount of data to be transmitted, what is the probability that the amount of energy required for its successful transmission is greater than a threshold level? Given the amount of available energy at transmitter, what is the largest amount of data that can be transmitted over the wireless channel reliably? The answers to these questions will provide the valuable design guidelines for the energy-efficient transmission of small data. In the following, we introduce two  \emph{data-oriented} energy efficiency metrics to address these design questions.

\section{Minimum energy consumption}

The fundamental service requirement of green IoT applications is to reliably transmit a certain amount of data to its destination over a given channel in a highly energy-efficient manner. We define a data-oriented energy utilization metric, namely minimum energy consumption (MEC), as the minimum amount of energy required to transmit a certain amount of data over a wireless channel. Let $H$ denote the amount of data to be transmitted. The MEC will be a function of $H$, denoted by $E_\mathrm{min}(H)$. For a given $H$ value, MEC will vary with the transmission power, the channel bandwidth, the channel realization, and the adopted transmission strategy. To illustrate further, we consider the MEC analysis for two adaptive transmission strategies over a point-to-point wireless link. We assume that the channel introduce flat fading. The noise spectral density at the receiver over the channel bandwidth $B$ is $N_0$, which leads a noise power of $N_0B$. 

\subsection{Continuous rate adaptation}

We first consider the continuous rate adaptation with constant power (CRA) transmission strategy.  Specifically, the transmitter adapts the transmission rate with the channel condition while maintaining constant transmission power $P_t$ \cite{ADD}. For the small data scenario, where $H$ is relatively small, data transmission will typically complete in a channel coherence time. Applying the Shannon capacity formula, the maximum instantaneous data rate for reliable transmission is equal to $B\cdot \log_2(1+P_tg/N_0B)$, where $g$ is the instantaneous channel power gain. The minimum time duration to finish data transmission is determined as $H/(B \log_2(1+P_tg/N_0B))$. We can the calculate MEC as the product of the transmission power and the minimum transmission time as $E_\mathrm{min}(H)=P_tH/(B \log_2(1+P_tg/N_0B))$, which varies with the instantaneous channel gain $g$. 

To address the earlier design questions, we define the energy outage rate (EOR) as the probability that MEC for a certain amount of data is greater than a threshold energy amount. In particular, EOR is mathematically defined as
 $\mathrm{EOR} = \Pr[E_\mathrm{min}(H) > E_\mathrm{th}]$, where $E_\mathrm{th}$ denotes the energy threshold. Equivalently, EOR can be calculated as the probability that the per-bit energy consumption is greater than a threshold value $E_\mathrm{th}/H$. The EOR for data transmission with CRA within a channel coherence time can be calculated as
\begin{equation}
\mathrm{EOR}^\mathrm{cra}
= F_g\left[\frac{N_0}{P_t/B}\left(\exp\bigr(\frac{\ln(2) H P_t}{B E_\mathrm{th}}\bigr)-1\right)\right],
\end{equation}
where $F_g(\cdot)$ denotes the CDF of the channel power gain $g$. 
 As such, EOR serves as an statistical characterization for the energy efficiency experienced by individual data transmission session with CRA. 

\begin{figure}[t]
\begin{center}
\includegraphics[height=10.2cm]{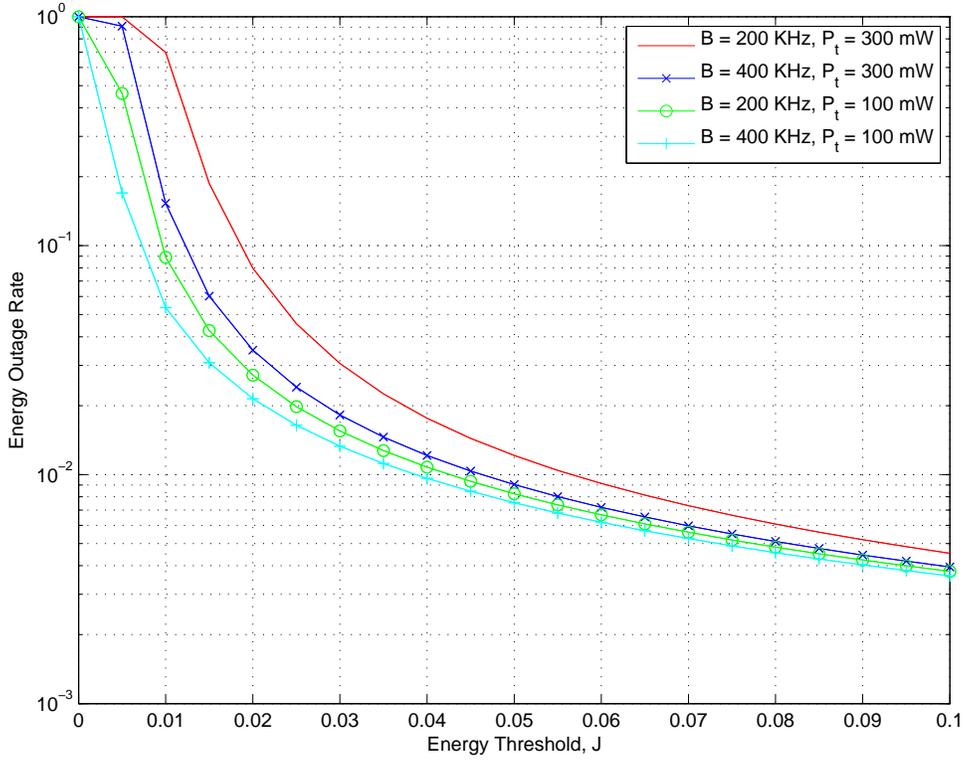}
\end{center}
\caption{Energy outage rate of CRA over slow Rayleigh fading channel ($H =$ 50 kB, and $\overline{g} =$ -10 dB).}
\label{DORHdiff}
\end{figure}

Fig. \ref{DORHdiff} plots the EOR of CRA transmission over slow Rayleigh fading channel as the function of the energy threshold $E_\mathrm{th}$ for different transmission parameter settings. We set the data amount to 50 kB and the average channel power gain to -10 dB. We can see that the EOR for all cases decreases with the energy threshold. Larger channel bandwidth help reduce the EOR for the same transmission power level, as expected by intuition. Meanwhile, for the same channel bandwidth, larger transmission power leads to larger EOR. Typically, larger transmission power help improve the received SNR for the same channel realization, which allows for higher transmission rate with CRA and in turn reduces the time duration to finishing data transmission. The transmission time reduction is, however, in logarithm with respect to $P_t$. As such, the MEC increases with $P_t$, which leads to high EOR. 

\subsection{Continuous power adaptation}

Power adaptation is a popular adaptive transmission strategy. Here, we consider the continuous power adaptation with constant rate (CPA) transmission strategy. In particular, the transmitter adapts the transmission power with the channel condition while maintain a constant received SNR, denoted by $\gamma_c$, under the peak power constraint $P_\mathrm{max}$ (also known as truncated channel inversion \cite{ADD}). Mathematically speaking, the transmission power $P_t$ is set to $\gamma_cN_0B/g$ when $g\ge g_T = \gamma_cN_0B/P_\mathrm{max}$, and 0 otherwise. Such transmission strategy can support error free transmission at rate $B\log_2(1+\gamma_c)$ when $g\ge g_T$. The MEC with CPA transmission can be calculated as 
\begin{equation}
E_\mathrm{min}(H) = \frac{\gamma_cN_0}{g}\frac{H}{\log_2(1+\gamma_c)},
\end{equation}
when $g\ge g_T$. We can see that MEC is inverse proportional to the channel gain $g$ for CPA, whereas for CRA, MEC is approximately proportional to $1/\log_2(g)$. Power adaptation can achieve better energy efficiency than rate adaptation at the cost of a certain probability of transmission outage. Note that when  $g< g_T$, the transmitter with CPA will hold the transmission until the channel condition improves, which may cause long delay. 

The EOR of CPA transmission can be calculated as the probability that $E_\mathrm{min}(H)$ given in (2) is greater than the energy threshold $E_\mathrm{th}$. Noting that the transmission will be held when  $g< g_T$, and as such, no transmission energy is consumed, the EOR for a certain amount of data with CPA can be evaluated as
\begin{equation}
\mathrm{EOR}^\mathrm{cpa}
= \left[F_g\left(\frac{\gamma_cN_0}{E_\mathrm{th}} \frac{H}{ \log_2(1+\gamma_c)}\right)- F_g(g_T)\right]/(1-F_g(g_T)).
\end{equation}

\begin{figure}[t]
\begin{center}
\includegraphics[height=10.2cm]{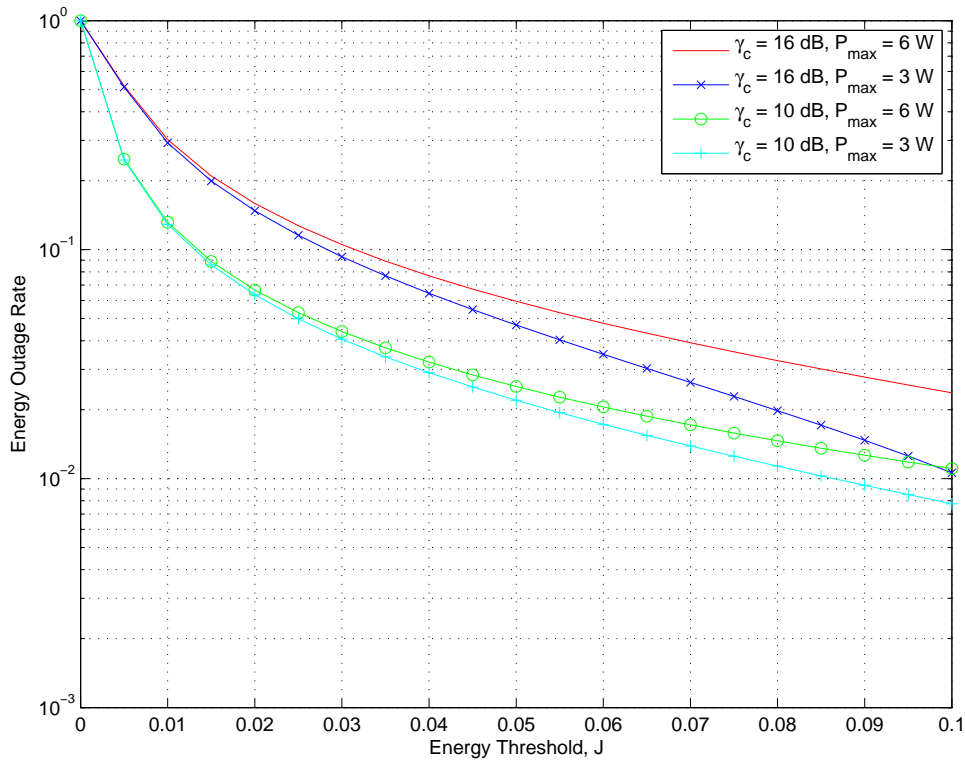}
\end{center}
\caption{Energy outage rate of CPA over slow Rayleigh fading channel ($H =$ 50 kB, $B = $ 200 kHz,  and $\overline{g} =$ -10 dB).}
\label{DORgbardiff}
\end{figure}

Fig. \ref{DORgbardiff} illustrates the EOR performance of CPA  over slow Rayleigh fading channels. In particular, we examine the effect of peak transmission power and target received SNR during transmission. We can see that maintaining a higher target received SNR with CPA leads to larger EOR. This can be explained by noting from Eq. (2) that the MEC with CPA will increase with $\gamma_c$. Another way to appreciate this behavior is to note that higher $\gamma_c$ implies larger transmission power during transmission on average. We also observe from Fig. \ref{DORgbardiff} that larger peak transmission power results in larger EOR, especially when the energy threshold is large. With CPA, larger $P_\mathrm{max}$ will lead to larger probability of transmission for the same target SNR. 
While leading to longer delay, smaller $P_\mathrm{max}$ will ensure that the system transmit only over more favorable channel condition and as such reduce the energy consumption. We conclude that different $P_\mathrm{max}$ values lead to different tradeoffs between energy efficiency and transmission delay.

\section{Maximum information delivery}

Most IoT devices are running on stringent energy budget. Many devices will be powered by energy harvesting from the ambient environment. Therefore, the efficient utilization of the limited energy resource for data transmission is of critical importance for IoT devices. In this section, we characterize the energy efficiency of individual data transmission session from the information delivery perspective. In particular, we define maximum information delivery (MID) as the maximum amount of information that can be reliably transmitted with a given amount of energy. Such characterization would be instrumental to the energy provision design for IoT devices. Mathematically, we denote MID by $H_\mathrm{max}(E)$, which is a function of the available energy amount, denoted by $E$. Here, MID will depends on the channel bandwidth, the channel realization, and the adopted transmission strategy. Note that MID can be applied to evaluate the bits/joule metric as $H_\mathrm{max}(E)/E$. We illustrate the MID analysis again by considering CRA and CPA transmission strategies over a point-to-point link for small data transmission scenario.

\subsection{Continuous rate adaptation}

With CRA transmission, the transmitter can transmit continuously for $E/P_t$ time period, where $P_t$ is the transmit power. If the amount of energy $E$ is relatively small and $E/P_t$ is less than a channel coherence time $T_c$, then the MID of CRA can be calculated as $H_\mathrm{max}(E) = (E/P_t) B\log_2(1+P_tg/N_0B)$ bits. The bits/joule energy efficiency becomes $B \log_2(1+P_tg/N_0B)/P_t$, which is changing with the instantaneous channel gain $g$. In particular, $H_\mathrm{max}(E)$ is approximately proportional to $\log_2 (g)$ for large $g$. When $E$ is large and $E/P_t$ spans multiple $T_c$'s, the MID with CRA is determined as $H_\mathrm{max}(E) = \sum_{i=1}^{N} T_c B \log_2(1+P_tg_i/N_0B)$, where $N$ is the number of $T_c$'s and $g_i$ is the channel power gain during the $i$th $T_c$. Here we assumed block fading channel, where the channel gain remains constant for one $T_c$ and changes to an independent value afterwards. 

Since MID is generally varying with the channel realization, we define the information outage rate (IOR)  as the probability that MID for a given amount of energy $E$ is less than a threshold entropy value, denoted by $H_\mathrm{th}$. Mathematically, IOR is given by $\Pr[H_\mathrm{max}(E) < H_\mathrm{th}]$. 
Apparently, the IOR analysis requires the statistics of  $H_\mathrm{max}(E)$, which depends on the channel bandwidth, the channel statistics, and the adopted transmission strategy. For example, when $E$ is small and $E/P_t$ is less than $T_c$,  the IOR with CRA can be calculated as
 \begin{equation}
\mathrm{IOR}^\mathrm{cra}
= F_g\left[\frac{N_0}{P_t/B}\left(\exp({\frac{\ln(2) H_\mathrm{th} P_t}{EB}})-1\right)\right].
\end{equation}
For the scenario that $E/P_t$ involves multiple $T_c$'s, the IOR will be equal to the probability that MID is less than $H_\mathrm{th}$, the evaluation of which will requires the distribution of the sum of $N$ independent random variables.  Further investigation of IOR for CRA transmission will be an interesting topic for future research.

\begin{figure}[t]
\begin{center}
\includegraphics[height=10.2cm]{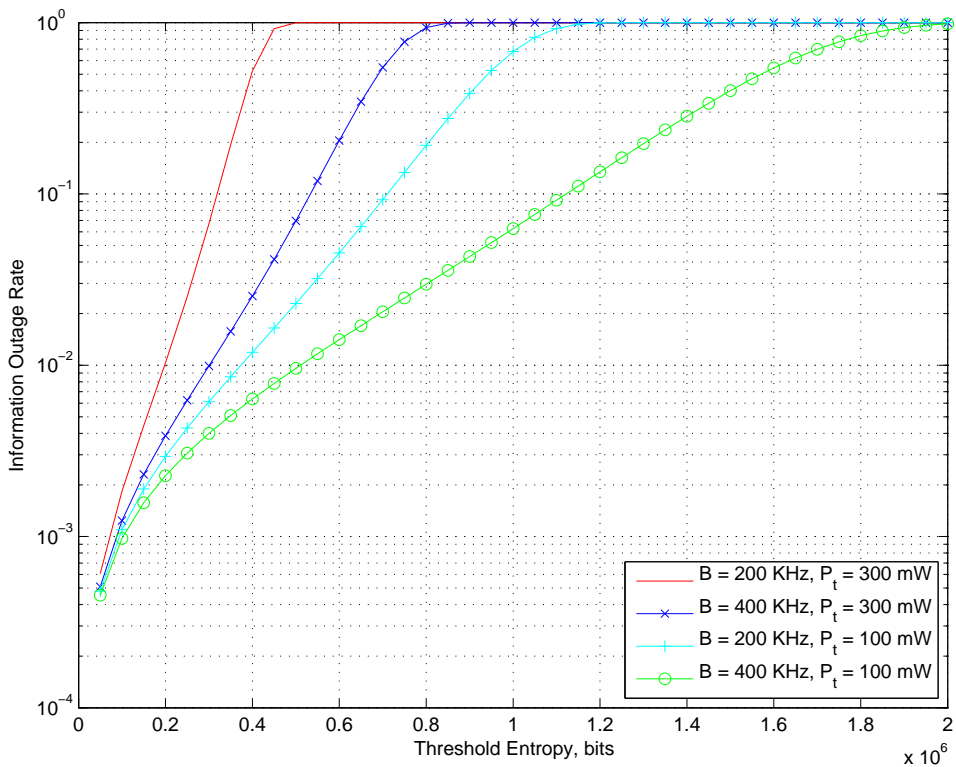}
\end{center}
\caption{Information outage rate of ORA over slow Nakagami fading channel ($E =$ 80 mJ, $m =$ 2, and $\overline{g} =$ -10 dB).}
\label{IORTdiff}
\end{figure}

Fig. \ref{IORTdiff} plots the IOR of CRA transmission as the function of threshold entropy  $H_\mathrm{th}$ for different transmission parameter settings over slow Nakagami fading channel. The amount of energy available for transmission usage is 80 mJ, which we assume can only support a transmission duration of one $T_c$. We can see that the IOR for all cases decrease with the threshold entropy. Larger channel bandwidth helps reduce the IOR for the same transmission power level, as expected by intuition. On the other hand, similar to EOR performance, IOR increases with larger transmission power for the same channel bandwidth. This is due to the fact that the transmission time is reducing linearly with transmission power whereas the transmission rate is increasing in logarithm with respect to $P_t$.

\subsection{Continuous power adaptation}

We now consider the MID analysis for CPA transmission strategy. Specifically, the transmit power is adaptively set to maintain a constant receive SNR of $\gamma_c$ while satisfying the peak power constraint  $P_\mathrm{max}$. As such, the transmission rate is fixed at $B\log_2(1+\gamma_c)$ with transmit power $\gamma_cN_0B/g$ when $g\ge g_T$ and equal to zero otherwise. Assuming slow fading environment where $E$ can only  support transmission over one channel coherence time, i.e. $E g/(\gamma_cN_0B)<T_c$, the MID can be calculated as 
\begin{equation}
H_\mathrm{max}(E) = \frac{Eg}{\gamma_cN_0}\log_2(1+\gamma_c). 
\end{equation}
We can see that the MID with CPA is linearly increasing with channel power gain $g$. Essentially, CPA transmission achieves higher energy efficiency than CRA at the cost of a certain probability of transmission outage. 

The IOR with CPA transmission can be calculated as the probability that $H_\mathrm{min}(E)$ is less than $H_\mathrm{th}$. Noting that no transmission power will be consumed over a coherence time if the channel power gain $g$ is less than $g_T$, the IOR for a certain amount of energy with CPA can be evaluated as
\begin{equation}
\mathrm{IOR}^\mathrm{cpa}
= \left[F_g\left(\frac{\gamma_cN_0}{E} \frac{H_\mathrm{th}}{\log_2(1+\gamma_c)}\right) - F_g(g_T)\right]/(1-F_g(g_T)).
\end{equation}

\begin{figure}[t]
\begin{center}
\includegraphics[height=10.2cm]{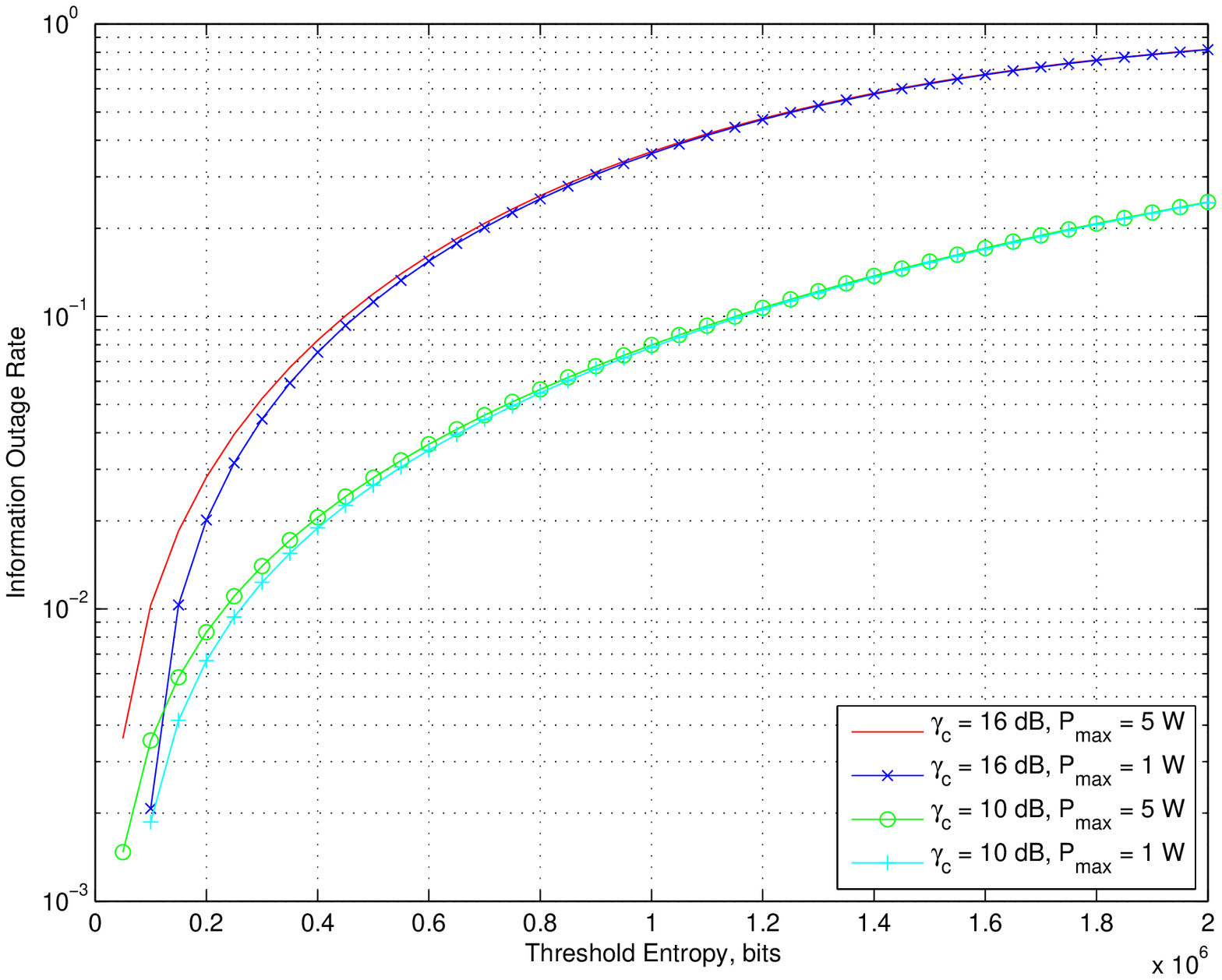}
\end{center}
\caption{Information outage rate of CPA over slow Nakagami fading channel l ($E =$ 80 mJ, $B = $ 200 kHz, $m =$ 2,  and $\overline{g} =$ -10 dB).}
\label{IORgbardiff}
\end{figure}

Fig. \ref{IORgbardiff} illustrates the IOR performance of CPA  over slow Nakagami fading channels. We again examine the effects of peak transmission power and target received SNR during transmission. We can see that maintaining a higher target received SNR with CPA leads to larger information outage rate. This behavior can be explained by noting that higher $\gamma_c$ implies larger transmission power during transmission on average. We also observe from Fig. \ref{IORgbardiff} that the peak transmission power level has minimum effect on IOR performance unless the entropy threshold is very small. Similar to EOR performance, the IOR performance degrades slightly when $P_\mathrm{max}$ increases. Smaller $P_\mathrm{max}$ will ensure that the system transmits only over more favorable channel condition and reduce the transmission power consumption on average. 

\section{Further considerations}

The above proposed data-oriented metrics characterize the energy efficiency performance of individual data transmission sessions over fading wireless channels. In particular, MEC prescribes the smallest amount of energy required for transmitting a certain amount of data over fading channels, whereas MID signifies the largest amount information that can be transmitted with a given amount of energy. Given the time-varying nature of wireless fading channels, these performance limits are described in a statistical sense, in terms of EOR and IOR, respectively. By specifying the best possible performance for individual transmission session, these limits will provide valuable guidelines to the development of practical energy-efficient transmission strategies for IoT applications. 

In previous sections, we illustrate the energy efficiency analysis of continuous rate and continuous power adaptive transmission strategies based on MEC and MID metrics. Both transmission strategies assume a certain channel state information (CSI) at the transmitter. The energy consumption associated with CSI acquisition was neglected in the analysis. When the amount of data is small, as in the `small data' scenario for IoT applications, the extra energy needed for CSI provision at the transmitter may be comparable to the transmit energy consumption. Further analysis on the overall energy consumption at the transmitter will be instrumental, especially for the comparison with transmission strategies requiring no CSI at the transmitter. 

Adaptive modulation and coding (AMC)  and automatic repeat request (ARQ) are two practical rate-adaptive transmission strategies that explore limited feedback from the receiver. AMC adapts the transmission rate for a certain reliability requirement whereas ARQ enhance the reliability with retransmission\cite{ADD}. With the proposed data-oriented energy efficiency metrics, we can compare the energy efficiency of AMC and AQR on the common ground of energy consumption per transmission session. Such study will generate new design insights on energy efficient transmission strategies for the limited CSI at the transmitter scenario. 

The power consumption at the transmitter includes transmission power and circuit power. The circuit power consumption is typically negligible compared with transmission power for conventional high power wireless transmission over long distance scenarios. Meanwhile, many IoT devices can not afford high transmission power. In such scenarios, the circuit power may even dominates the overall power consumption \cite{1532220}. Furthermore, to maintain the same output power, the power consumption of RF amplifier may vary with the chosen modulation scheme, as different modulation scheme will lead to different RF amplifying efficiency \cite{Lee_CMOS}. As such, the energy efficiency analysis with these practical considerations will entail new challenges. 



Energy harvesting is an essential technology for green IoT and will provide IoT devices with eternal power supply. Meanwhile, the amount of energy that can be harvested over a certain time period varies considerably. The MID analysis together with the energy arrival process characterization will be essential to the successful design of the energy-aware scheduling algorithms. The general design goal is to ensure that the IoT devices will have sufficient energy to complete their transmission with high probability. With the data-oriented energy consumption analysis, we can analyze and compare the performance of different scheduling algorithms for diverse target applications. 


\section{Concluding Remarks}

In this article, we present a novel data-oriented approach for energy efficiency charaterization of wireless transmission systems. We target at the small data transmission scenario for IoT applications. In particular, we introduce two data-oriented performance limits on energy efficiency for arbitrary wireless data transmission. As their initial application, we analyze two channel adaptive transmission strategies and examine the effects of system parameters on their energy efficiency performance. We observe that the data-oriented approach can bring interesting new insights on green wireless transmission over fading channels. 
This article serves as an initial introduction to the data-oriented approach for green wireless transmission design. There are many important aspects to be addressed, including the limited and no CSI at transmitter scenarios. We expect that the data-oriented perspective will stimulate promising novel design of green wireless transmission strategies for IoT applications. 

\bibliographystyle{IEEEtran}
\bibliography{IEEEabrv,IEEETED}

\end{document}